\documentclass[a4paper,twocolumn]{article}

\usepackage[colorlinks=true,linkcolor=blue,urlcolor=blue,citecolor=blue]{hyperref}
\usepackage[affil-it]{authblk}
\usepackage{graphics}
\usepackage{gensymb}
\usepackage[font=scriptsize]{caption}
\usepackage[margin=0.7in]{geometry}

\begin{document}
		
\title{\textbf{Selective reflection from Rb layer with thickness below $\lambda$/12 and applications}}

\author[1]{A. Sargsyan}
\author[1]{A. Papoyan}
\author[2]{I.G. Hughes}
\author[2]{C.S. Adams}
\author[1]{D. Sarkisyan}

\affil[1]{\fontsize{10}{13}\selectfont Institute for Physical Research, NAS of Armenia, 0203, Ashtarak-2, Armenia}
\affil[2]{Joint Quantum Centre (JQC) Durham-Newcastle, Department of Physics, Durham University, South Road, Durham DH1 3LE, United Kingdom}

\twocolumn[
\begin{@twocolumnfalse}
\maketitle
\begin{abstract}
We have studied the peculiarities of selective reflection from Rb vapor cell with thickness $L <$ 70 nm, which is over an order of magnitude smaller than the resonant wavelength for Rb atomic D$_1$ line $\lambda$ = 795 nm. A huge ($\approx$ 240 MHz) red shift and spectral broadening of reflection signal is recorded for $L =$ 40 nm caused by the atom-surface interaction. Also completely frequency resolved hyperfine Paschen-Back splitting of atomic transitions to four components for $^{87}$Rb and six components for $^{85}$Rb is recorded in strong magnetic field ($B >$ 2 kG).
\bigskip
\end{abstract}
\end{@twocolumnfalse}
]
\maketitle

Selective reflection (SR) of resonant laser radiation from the interface between an atomic vapor and the dielectric window of spectroscopic cells initially observed by R. Wood in 1909 \cite{wood1} has revisited in 1970s when narrow-linewidth tunable cw lasers became available \cite{woerdman1,nienhuis1,weis1}. Thanks to high contrast sub-Doppler signal response, the SR technique became a powerful spectroscopic tool, which was successfully used, in particular, to study the van der Waals (vdW) interaction of atoms with dielectric window of the cell manifested by a red shift of the SR frequency \cite{failache1,hamdi1,bloch1,fichet1}. Spectroscopy with atomic vapor nanometric-thickness cells (NC) is another technique capable of yielding information on atom-surface processes. The vdW interaction of Rb and Cs atoms confined in the NC was studied in \cite{whittaker1} using resonant fluorescence spectra.

Merging the SR and NC methods significantly extends their applied interest. On account of the dispersive shape of the signal, the SR from NC was used to lock the laser's frequency \cite{gazazyan1}. Recently it was demonstrated that the SR from the Rb vapor NC with $L \approx$ 370 nm thickness is convenient to form frequency reference of atomic transitions \cite{sargsyan1}. It was also shown that the same technique implemented for D$_1$ line of Cs and 300 nm gap thickness is expedient to study the behavior of atomic transitions in strong magnetic field \cite{sargsyan2}.

In this Letter we focus on the peculiarities of selective reflection for the conditions when the response of a resonant medium is formed over an essentially sub-wavelength depth in an atomic vapor layer geometrically restricted by the two inner surfaces of the nanometric-thickness cell ($L<$ 70 nm).

\begin{figure}[h!]
	\centering
	\resizebox{0.43\textwidth}{!}{\includegraphics{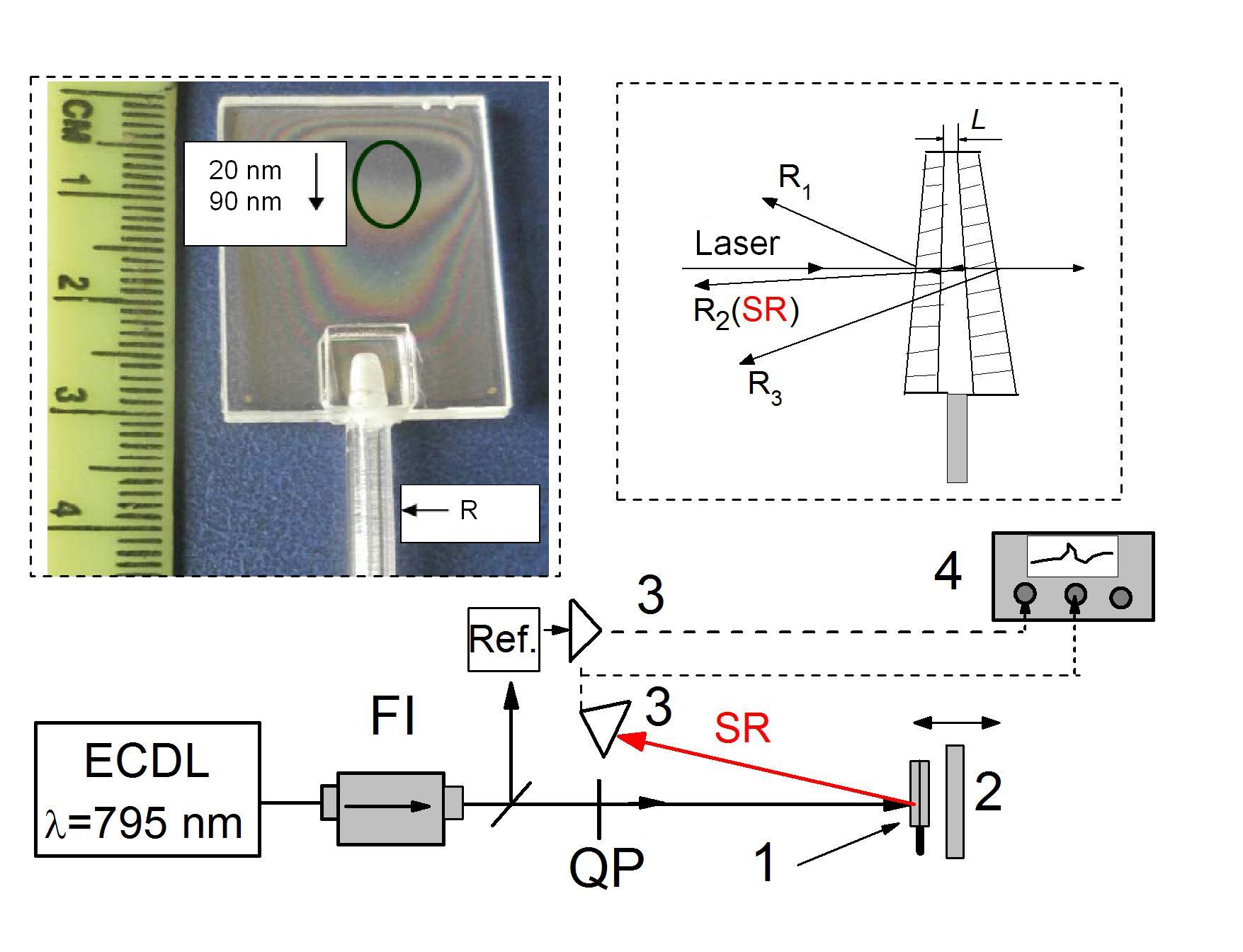}
	}
	\caption{Layout of the experimental setup: (ECDL) cw laser, (FI) Faraday isolator, ($1$) Rb NC inside an oven, ($2$) permanent magnet, ($3$) photodetector, (Ref.) reference spectrum unit, ($4$) oscilloscope, (QP) quarter-wave plate. Upper left inset: photograph of the NC, oval marks 20 -- 90 nm region. Upper right inset: geometry of the 3 reflected beams. The beam (SR) propagates in the direction of R$_2$.}
	\label{fig:Fig1}
\end{figure}

The experimental arrangement is schematically sketched in Fig.\ref{fig:Fig1}. The collimated $P_L \sim$ 0.5 mW circularly-polarized beam of a single-frequency narrow-band ($\gamma_L$ $\sim$ 1 MHz) cw external cavity diode laser with $\lambda=$ 795 nm was directed at normal incidence onto the specially fabricated Rb NC with a large aperture area of smoothly changing thickness in the rangle $L=$ 20 -- 90 nm (see the upper left inset). The SR beam was carefully separated from two other beams reflected from the front and rear surfaces of the cell (see the upper right inset), and after passing through a $\lambda =$ 795 nm interference filter was recorded by a photodiode (3).

A fraction of laser radiation was branched to the frequency reference unit (Ref.) with an auxiliary $L=\lambda$ Rb NC, with a transmission spectrum exhibiting narrow velocity selective optical pumping (VSOP) resonances located exactly at atomic transition frequencies \cite{sargsyan3}. The reference signal was recorded simultaneously with the SR signal by a two-channel digital oscilloscope while linearly scanning the laser frequency across the D$_1$ resonance. The scanning rate was chosen to be slow enough for assuring establishment of a steady-state interaction regime.

The necessary vapor density ($\approx$ 8$\times$10$^{13}$ cm$^3$) was attained by heating the cell's thin sapphire reservoir (R) containing metallic Rb to $T_R \approx$ 150 \degree C while keeping the window temperature some 20 \degree C higher. Note that the NC can be heated up to 450 \degree C (see \cite{keaveney1,sargsyan3}). To study magnetic field-induced processes, a calibrated strong permanent neodymium magnet (2) was mounted on a micrometric-step translation stage in the proximity of the cell's rear window. The $B$-field strength was varied by simple longitudinal displacement of the magnet.  

\begin{figure}[h!]
	\centering
	\resizebox{0.48\textwidth}{!}{\includegraphics{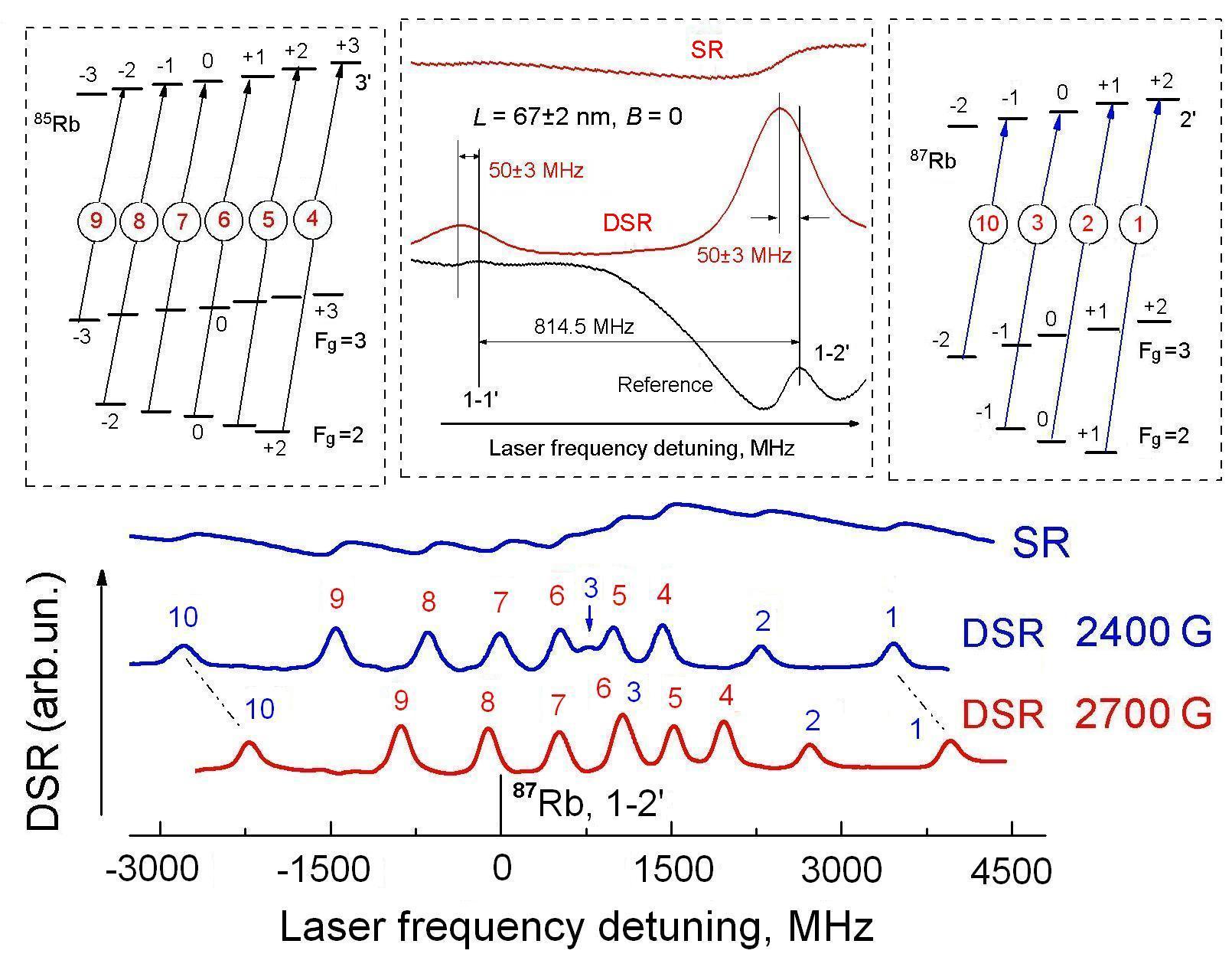}
	}
	\caption{Selective reflection spectrum (SR, upper line) and its derivative (DSR) for $L$ = 67 $\pm$2 nm, recorded at $B$ = 2.4 kG (middle curve), and 2.7 kG (lower curve). For labeling, see the upper left- and right-side insets presenting $\sigma^+$-polarized transition diagrams for $^{85}$Rb and $^
		{87}$Rb, respectively. The upper middle inset: $B$ = 0 spectra (from top to bottom: SR, DSR, and reference) exhibiting $\sim$ 50 MHz red shift with respect to $F_g=1 \rightarrow F_e=1,2$ transitions of $^{87}$Rb.}
	\label{fig:Fig2}
\end{figure}

The SR spectrum on the D$_1$ line $F_g=2,3 \rightarrow F_e=3$ transitions of $^{85}$Rb and $F_g=1,2 \rightarrow F_e=2$ transitions of $^{85}$Rb\footnote{In the figures the excited levels are marked by primes.} for the cell thickness of 67 nm and longitudinal magnetic field $B \approx$ 2.4 kG is shown in Fig.\ref{fig:Fig2} (upper spectrum). The lower lines are the derivatives of SR signal (DSR) formed in a real time by the Siglent oscilloscope \cite{sargsyan1}. The DSR peaks labeled 1–3 and 10 belong to $^{87}$Rb, and the DSR peaks 4–9 belong to $^{85}$Rb (see the upper diagrams). On account of the sub-Doppler linewidth of DSR ($\gamma_{DSR}\approx$ 200 MHz FWHM), all ten transition components for $B$ = 2.4 kG are spectrally well resolved. For $B$ = 2.7 kG, complete frequency resolution is observed for all transition components except overlapping peaks 3 and 6. The influence of vdW interaction between the atoms and dielectric windows of the NC starts to feature in the spectra when $L$ < 100 nm \cite{hamdi1,bloch1}, manifested as a red shift of the SR signal frequency. This shift is clearly observable for $B$ = 0, $L\approx$ 67 nm, as is shown in the middle inset of Fig.\ref{fig:Fig2} (see figure caption).

\begin{figure}[b!]
	\centering
	\resizebox{0.47\textwidth}{!}{\includegraphics{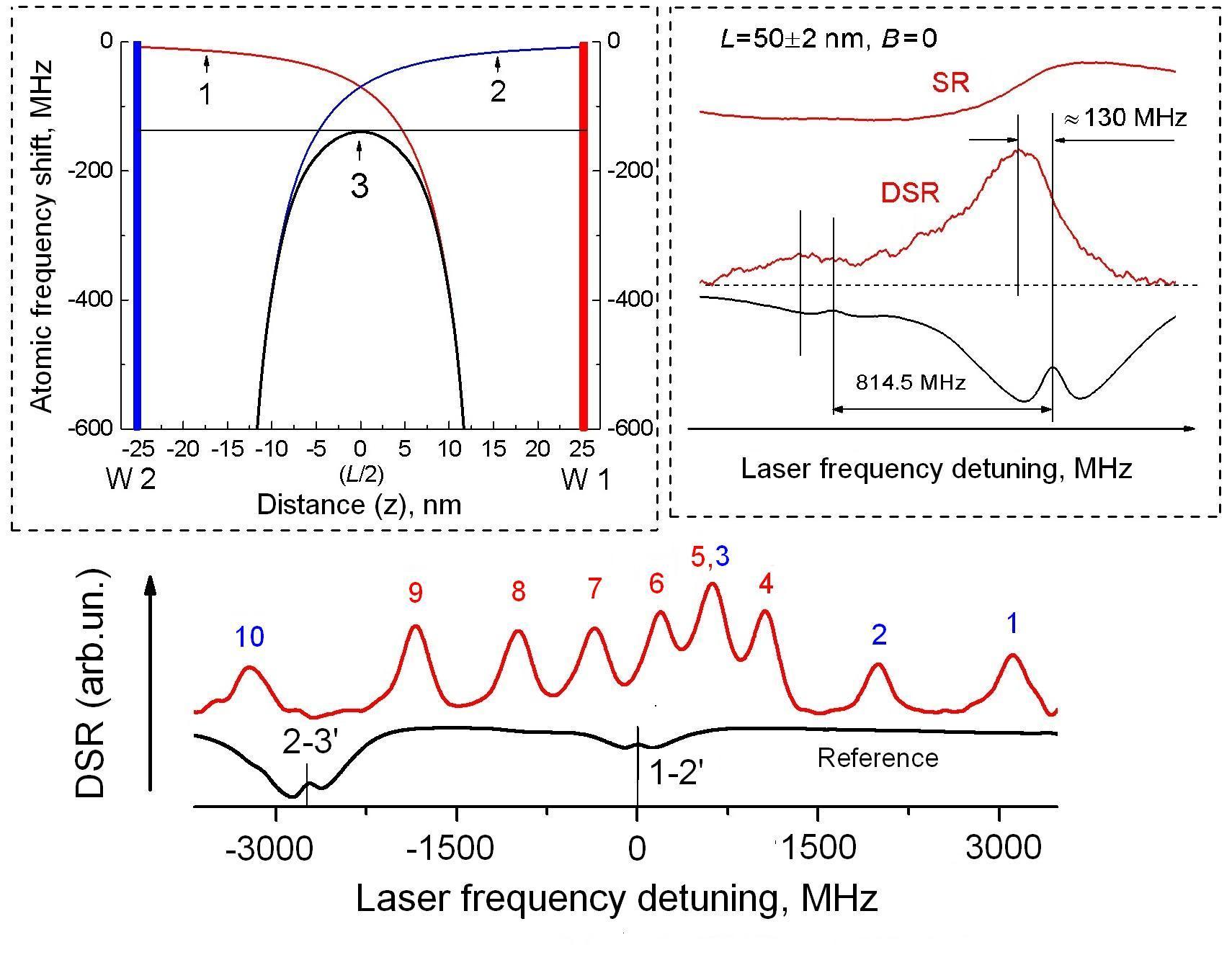}
	}
	\caption{Experimental DSR spectrum for $L$ = 50 $\pm$2 nm recorded at $B$ = 2.15 kG. Upper-right inset: $B$ = 0 spectra (from top to bottom: SR, DSR, and reference), exhibiting $-$130 MHz red shift with respect to to $^{87}$Rb $F_g=1 \rightarrow F_e=1,2$ reference frequencies. Upper-left inset: the estimate of distance-dependent vdW red shift; the horizontal line tangent to an overall curve 3 marks the minimum shift value ($-$130 MHz).}
	\label{fig:Fig3}
\end{figure}

The influence of the applied magnetic field on the hyperfine structure is characterized by parameter $B_0=A_{HFS}/\mu_B$, where $A_{HFS}$ is the hyperfine coupling constant for 5S level, and $\mu_B$ is the Bohr magneton \cite{zentile1}. $B_0 \approx$ 0.7 kG for $^{85}$Rb, and $B_0 \approx$ 2 kG for $^{87}$Rb. When $B<B_0$ (Zeeman regime), the splitting of levels is described by the total angular momentum of the atom $\textbf{F}$ = $\textbf{J}$ + $\textbf{I}$ and its projection $m_F$, where \textbf{J} is the total electron angular momentum, and \textbf{I} is the nuclear spin angular momentum. The $F$ and $m_F$ notation is used in the inset diagram of Fig.\ref{fig:Fig2}. Decoupling between \textbf{J} and \textbf{I} develops when $B\ge B_0$. $F$ is no longer a good quantum number, and the splitting of atomic levels is described by the projections $m_J$ and $m_I$ (hyperfine Paschen-Back regime) \cite{olsen1,sargsyan4,weller1}. In this regime four and six atomic transitions belonging to $^{87}$Rb and $^{85}$Rb, respectively, remain in the spectra (as is evident in Fig.\ref{fig:Fig2} and Fig.\ref{fig:Fig3}). 

The recorded DSR spectrum for $L$ = 50 $\pm$2 nm and $B \approx$ 2.15 kG is shown in Fig.\ref{fig:Fig3}. For this measurement, the reservoir temperature was somewhat increased ($T_R$ = 165 \degree C) to compensate for the thickness reduction. In spite of some broadening of DSR ($\gamma_{DSR} \approx$ 250 MHz), all the peaks are well resolved, except 3 and 5, which are overlapped. The inset in the upper-right corner shows the $B$ = 0 spectra, which clearly indicate the increase of vdW red shift to $-$130 MHz with respect to unperturbed atomic transitions. To estimate the frequency shift for $^{87}$Rb, D$_1$ line arising from interaction of an individual atom with two dielectric windows of the cell (w1 and w2), we have plotted in the upper-left inset the distance-dependent frequency shifts separately for w1: $\Delta\nu_{vdW} = -C_3/z_1^3$ (curve 1), and for w2: $\Delta\nu_{vdW} = -C_3/z_2^3$ (curve 2), where $z_1$ and $z_2$ are distances of the Rb atom from w1 and w2, respectively (in $\mu$m). The overall shift is depicted by curve 3, which is the sum of curves 1 and 2. For $L$ = 50 nm, the recorded $-$130 MHz shift is obtained by taking vdW coefficient $C_3$ = 1.0$\pm$0.1 kHz$\times\mu$m$^3$, which is consistent with the results of previous studies \cite{bloch1}. As is seen from the diagram, the maximum of the spectral density of an overall signal (the shifted peak) corresponds to the minimum value of the shift, so this value is estimated by the expression

\begin{equation}
\label{eq::1}
\Delta\nu_{vdW} = -2C_3/(L/2)^3. 
\end{equation}

\begin{figure}[h!]
	\centering
	\resizebox{0.43\textwidth}{!}{\includegraphics{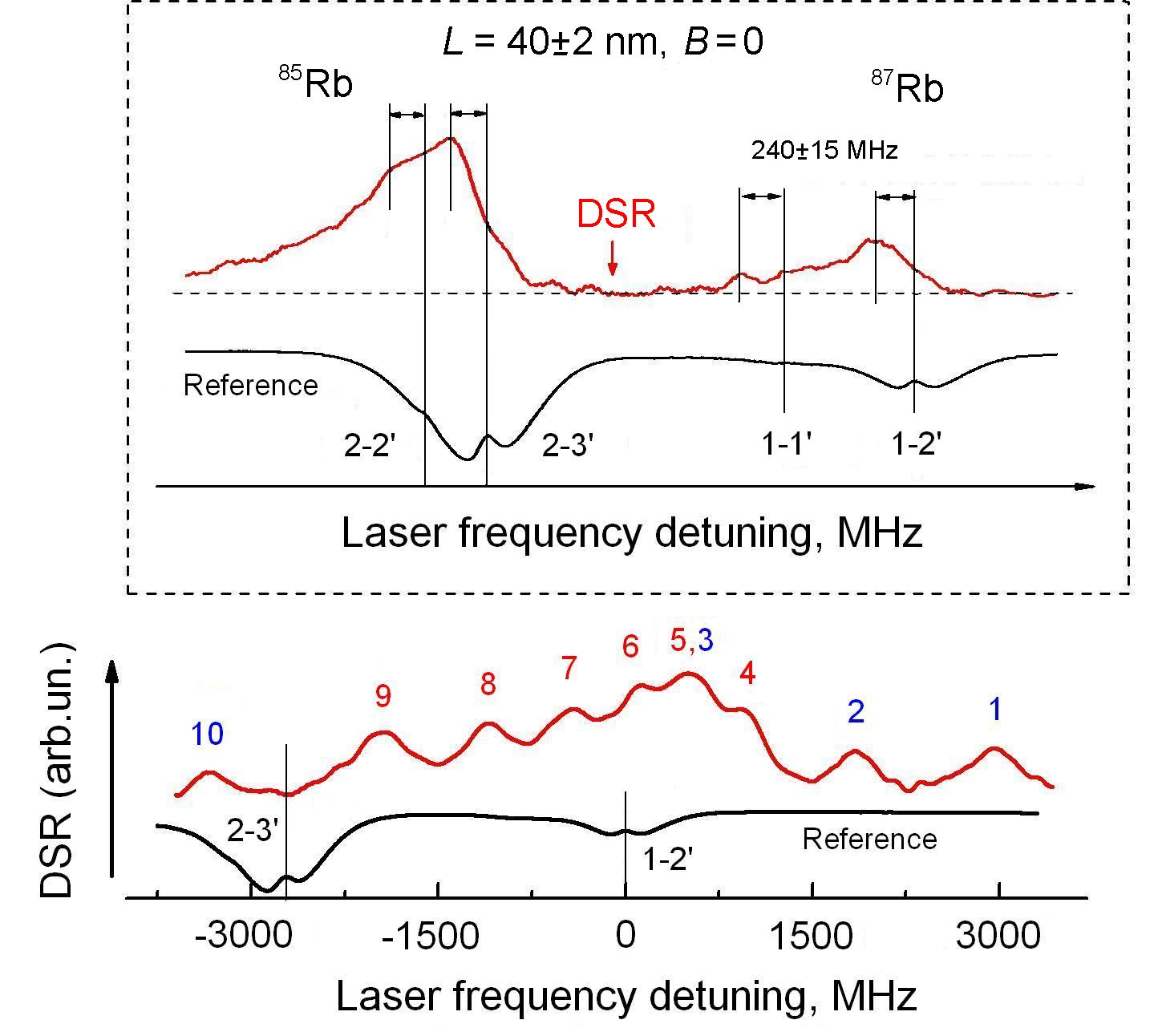}
	}
	\caption{Experimental DSR spectrum for $L$ = 40 $\pm$2 nm recorded at $B$ = 2.15 kG (upper trace) and $B$ = 0 reference spectrum (lower trace). The inset:  DSR and reference spectra at $B$ = 0 showing $-$240$\pm$15 MHz red shift with respect to $^{87}$Rb $F_g=1 \rightarrow F_e=1,2$ and $^{85}$Rb $F_g=2 \rightarrow F_e=2,3$ reference frequencies.}
	\label{fig:Fig4}
\end{figure}

To record SR signals above the noise level for even smaller cell thickness $L$ = 40$\pm$2 nm, the cell temperature was further increased to $T_R$ = 180 \degree C (vapor density $\sim 5\times10^{14}$ cm$^{-3}$). The DSR spectrum for this thickness and $B \approx$ 2.15 kG is presented in Fig.\ref{fig:Fig4}. In these conditions the transition peaks become broader, but the linewidth $\gamma_{DSR} \approx$ 380 MHz still does not exceed $\sim$ 500 MHz transition width obtained with 1 mm-long Rb cell \cite{weller1}. An estimation of the number of atoms contributing to SR signal (Fig.\ref{fig:Fig4}) for 0.5 mm beam diameter, $L$ = 40 nm, and atomic density 5$\times$10$^{14}$ cm$^{-3}$ gives $\sim 5\times10^6$. Note that the vdW shift for $^{85}$Rb and $^{87}$Rb transitions for the same $L$ value is approximately equal.

\begin{figure}[b!]
	\centering
	\resizebox{0.4\textwidth}{!}{\includegraphics{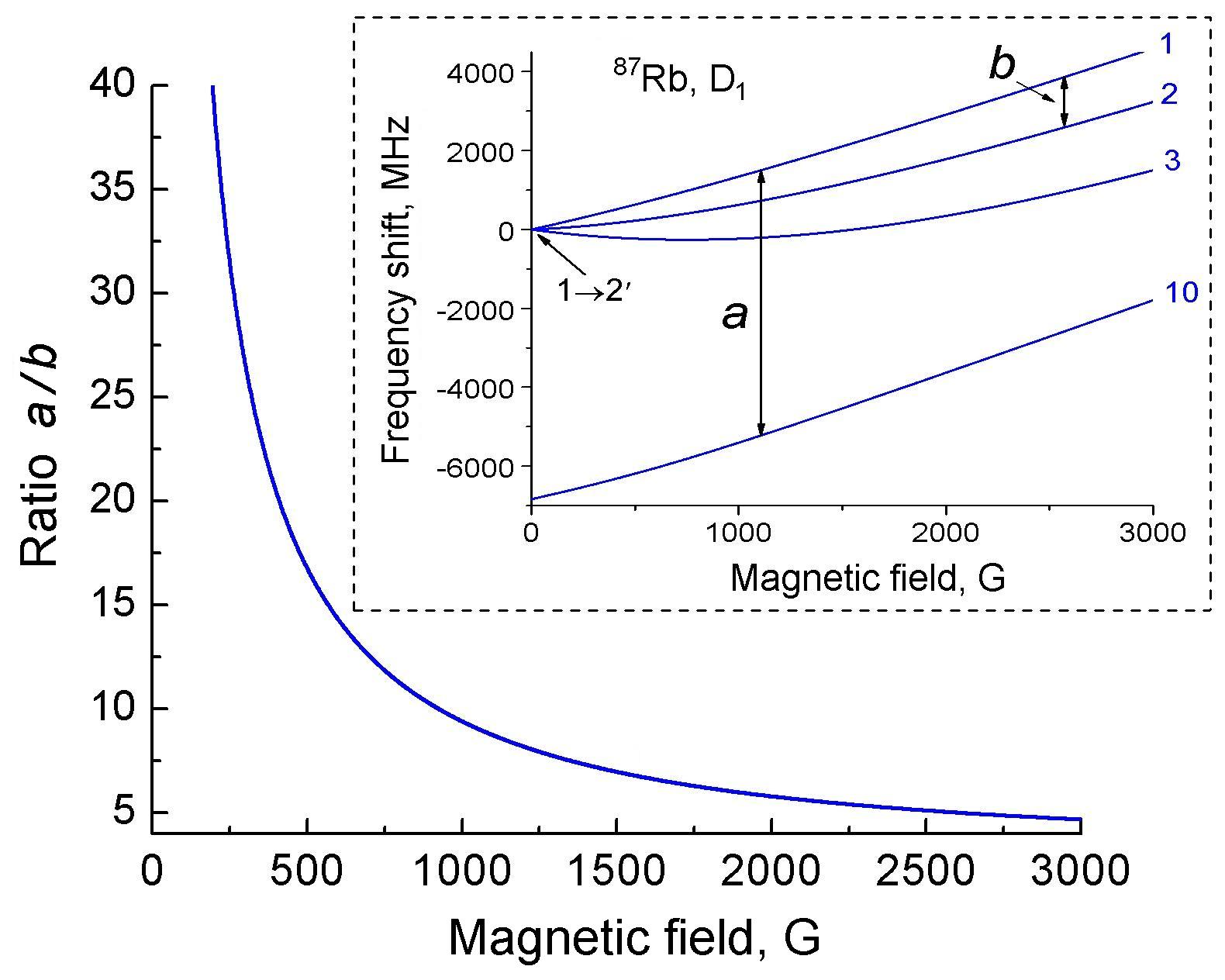}
	}
	\caption{$B$-field dependences of DSR frequencies exploited for realization of two ways to determine the magnetic field value (see text).}
	\label{fig:Fig5}
\end{figure}

Complete resolution of all the DSR peaks makes favorable the determination of a $B$-field, which can be done in two convenient ways: a) by measuring the frequency shift of DSR peak 1 ($\nu_1$) from the reference transition $F_g=1 \rightarrow F_e=2$ (see the inset of Fig.\ref{fig:Fig5}), also taking into account the value of vdW shift; b) by measuring the frequency separations between the individual transitions $a=\nu_{10}-\nu_1$ and $b=\nu_2-\nu_1$, and exploiting the dependence of the $a/b$ ratio on the $B$-field. Note that for case b) there is no need for a frequency reference, and also the value of vdW shift is not important, being the same for all four transitions. Both $B$-field measurement techniques are graphically elucidated in Fig.\ref{fig:Fig5}.  

\begin{figure}[t!]
	\centering
	\resizebox{0.4\textwidth}{!}{\includegraphics{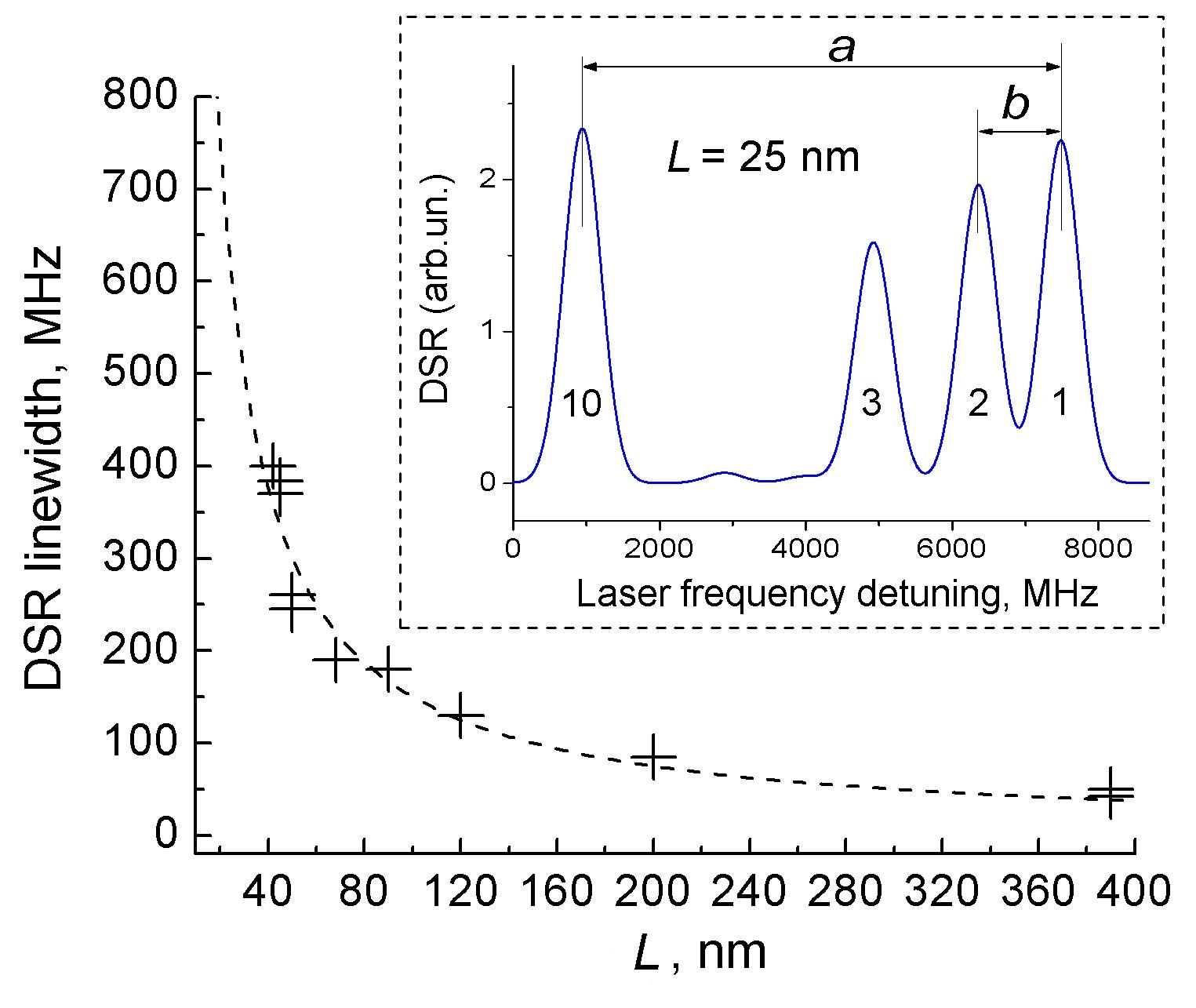}
	}
	\caption{Dependence of DSR linewidth $\gamma_{DSR}$ versus $L$. The inset: four DSR peaks of the $^{87}$Rb D$_1$ line at $L$ = 25 nm and $B$ = 2 kG; $\gamma_{DSR} \approx$ 600 MHz. The inset shows theoretical spectrum calculated using dependences of the frequency shifts and transition probabilities versus magnetic field, see \cite{sargsyan2}.}
	\label{fig:Fig6}
\end{figure}

The SR and DSR broadening observed with reduction of $L$ and/or increase of $T_R$ is caused by vdW interaction, as well as atom-window and atom-atom collisions. The measured DSR broadening $\gamma_{DSR}$ versus $L$ (cross marks) is presented in Fig.\ref{fig:Fig6}, along with fitted $\gamma_{DSR}$[MHz] $\approx$ 15000$/L$[nm] dependence (dotted line). Measuring magnetic field with nanometric-scale spatial resolution remains a challenge in a variety of problems, in particular when strong gradient fields are applied (e.g. 40 kG/mm in \cite{folman1}). Even for $L$ = 25 nm, $\gamma_{DSR}$ in our experiment is 600 MHz, so measuring the ratio $a/b$ for a NC filled with $^{87}$Rb isotope, it is possible to determine the $B$-field value with rather high precision. 

\begin{figure}[h!]
	\centering
	\resizebox{0.4\textwidth}{!}{\includegraphics{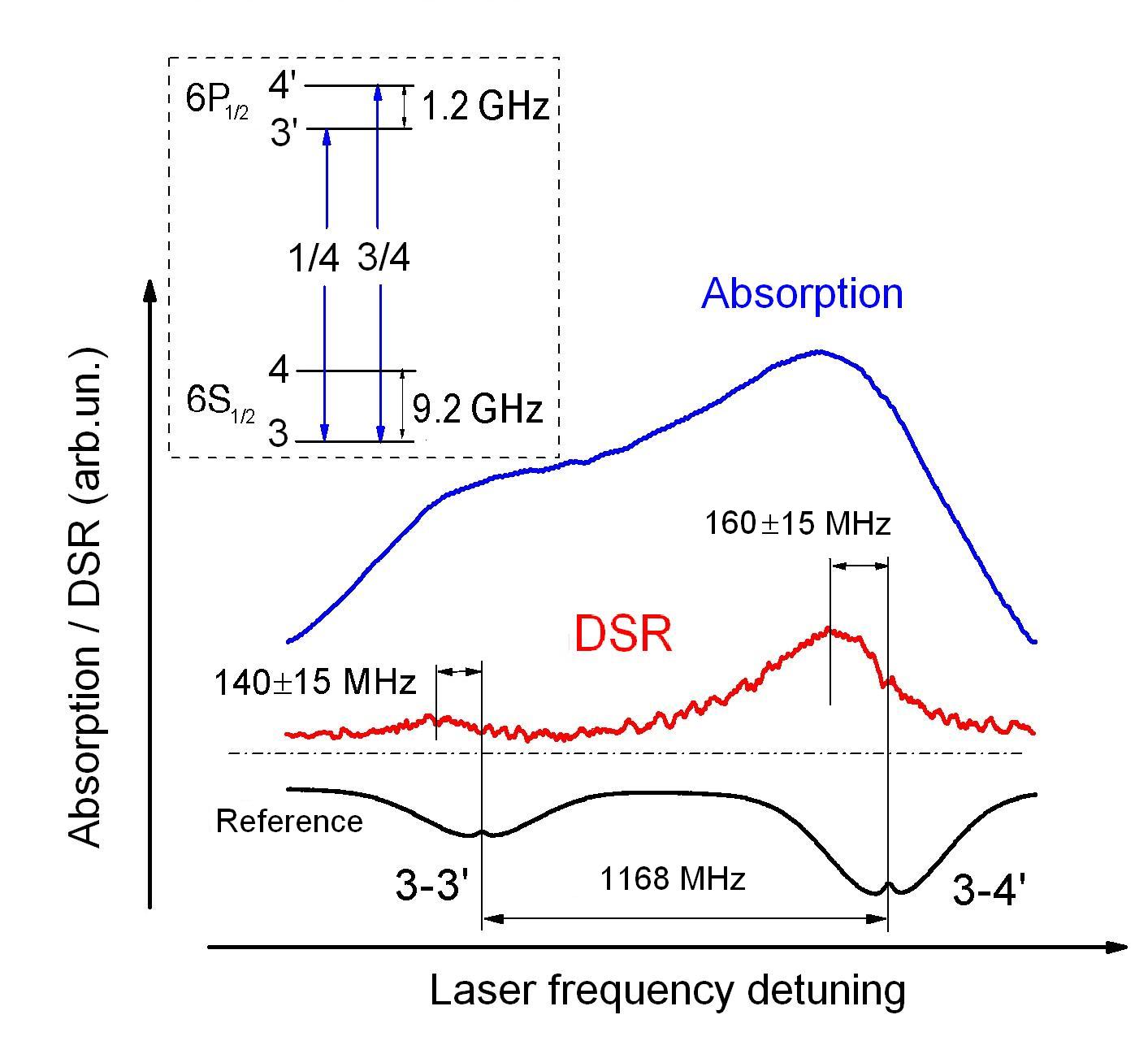}
	}
	\caption{Absorption spectrum (upper trace) and DSR spectrum (middle trace) recorded on Cs D$_1$ line and for $L \approx 52\pm3$ nm, $P_L \approx$ 0.2 mW, NC temperature $\approx$ 170 \degree C. The lower trace is the reference spectrum obtained with an auxiliary NC with $L=\lambda$, with VSOP resonances serving as $^{133}$Cs F$_g$=3 $\rightarrow$ F$_e$=3,4 transition frequency markers  \cite{sargsyan3}. Inset shows the diagram of these transitions with indication of their relative intensities.}
	\label{fig:Fig7}
\end{figure}

For comparison of the DSR with the absorption spectrum under the same conditions (for parameters, see the caption of Fig.\ref{fig:Fig7}), we have used $F_g=3 \rightarrow F_e=3.4$ transitions of Cs D$_1$ shown in the inset of Fig.\ref{fig:Fig7}, for which the frequency separation of excited levels, $\sim$ 1.2 GHz, is the largest among the alkali metals. The latter is advantageous for avoiding the overlap of absorption lines of neighboring transitions. In Fig.\ref{fig:Fig7} the upper curve shows the absorption spectrum for $L \approx$ 52 nm, and the middle curve presents DSR spectrum. The lowest line is the reference transmission spectrum obtained with an auxiliary NC with $L=\lambda$, exhibiting VSOP resonances at $F_g=3 \rightarrow F_e=3,4$ transitions of $^{133}$Cs \cite{sargsyan3}. The $F_g=3 \rightarrow F_e=3$ and $F_g=3 \rightarrow F_e=4$ transitions are strongly overlapped in the absorption spectrum, while DSR-peaks for the same transitions are obviously advantageous: they are background-free, completely resolved, and exhibit noticeable frequency red shift ($\sim$ 140 MHz and 160 MHz, respectively). Estimation of $C_3$ vdW coefficient obtained from Eq.(\ref{eq::1}) gives $C_3=1.4\pm0.1$ kHz$\times\mu$m$^3$. 

Owing to the low divergence and relatively high power of the SR signal ($\approx$ 0.5 $\%$ of the incident radiation), and its linear response remaining up to $P_L \sim$ 5 mW, and high signal-to-noise ratio, the DSR technique can be used for high-distance remote monitoring and mapping of both homogeneous and highly inhomogeneous $B$-fields in a wide range with $\sim$ 40 nm spatial resolution. The DSR technique employing nanometric-thickness cells is simple and easily realizable. We note that the recent development of a glass NC \cite{whittaker2} can make this technique widely available, both for studies of atom-surface interaction and $B$-field mapping.

\vspace{2 mm}The work was partially supported by MES RA (projects No. 15T-1C040 and 15T-1C277). The authors are grateful to A. Sarkisyan for fabrication of nanocell, and to A. Tonoyan for theoretical calculations.

\end{document}